\documentclass[preprint,english,german]{aastex63}
\usepackage{multirow}
\usepackage{graphicx}
\usepackage{mwe}
\usepackage[version=4]{mhchem}
\usepackage{xcolor}
\usepackage{xspace}
\usepackage{comment}
\usepackage{placeins}
\usepackage[T1]{fontenc}
\usepackage[utf8]{inputenc}

\received{--}
\revised{--}
\accepted{--}
\submitjournal{ApJ}

\shorttitle{A New Method for Simulating Photoprocesses}
\shortauthors{Mullikin et al.}
\graphicspath{{./}{figures/}}

\begin{document}

\title{A New Method for Simulating Photoprocesses in Astrochemical Models }

\correspondingauthor{Christopher N. Shingledecker}
\email{cshingledecker@benedictine.edu}

\author{Ella Mullikin}
\affil{Department of Chemistry, Wellesley College, Wellesley, MA 02481, USA}

\author{Hannah Anderson}
\affil{Department of Chemistry, Wellesley College, Wellesley, MA 02481, USA}

\author{Natalie O'Hern}
\affil{Department of Chemistry, Wellesley College, Wellesley, MA 02481, USA}

\author{Megan Farrah}
\affil{Department of Chemistry, Wellesley College, Wellesley, MA 02481, USA}

\author[0000-0002-0722-1151]{Christopher R. Arumainayagam}
\affil{Department of Chemistry, Wellesley College, Wellesley, MA 02481, USA}

\author[0000-0001-7591-1907]{Ewine F. van Dishoeck}
\affil{Leiden Observatory, Leiden University, P.O. Box 9513, NL-2300 RA Leiden, The Netherlands}
\affil{Max-Planck-Institut für extraterrestrische Physik, D-85748 Garching, Germany}

\author[0000-0002-9667-5904]{Perry A. Gerakines}
\affil{Astrochemistry Laboratory, NASA Goddard Space Flight Center, Greenbelt, MD 20771, USA}

\author[0000-0003-1684-3355]{Anton I. Vasyunin}
\affil{Ural Federal University,  Ekaterinburg, Russia}
\affil{Visiting Leading Researcher, Engineering Research Institute 'Ventspils International Radio
Astronomy Centre' of Ventspils University of Applied Sciences, 
 In\v{z}enieru 101, Ventspils LV-3601, Latvia}

\author[0000-0001-7031-8039]{Liton Majumdar}
\affil{School of Earth and Planetary Sciences, National Institute of Science Education and Research, HBNI, Jatni 752050, Odisha, India}

\author[0000-0003-1481-7911]{Paola Caselli}
\affil{Center for Astrochemical Studies
Max Planck Intitute for Extraterrestrial Physics
Garching, Germany}

\author[0000-0002-5171-7568]{Christopher N. Shingledecker}
\affil{Center for Astrochemical Studies
Max Planck Intitute for Extraterrestrial Physics
Garching, Germany}
\affil{Institute for Theoretical Chemistry 
University of Stuttgart 
Pfaffenwaldring 55, 70569 
Stuttgart, Germany}
\affil{Department of Physics \& Astronomy,
Benedictine College,
Atchison, KS 66002, USA}



\begin{abstract}

	We propose a new model for treating solid-phase photoprocesses in interstellar ice analogues. In this approach, photoionization and photoexcitation are included in more detail, and the production of electronically-excited (suprathermal) species is explicitly considered. In addition, we have included non-thermal, non-diffusive chemistry to account for the low-temperature characteristic of cold cores. As an initial test of our method, we have simulated two previous experimental studies involving the UV irradiation of pure solid O$_2$. In contrast to previous solid-state astrochemical model calculations which have used gas-phase photoabsorption cross-sections, we have employed solid-state cross-sections in our calculations. This method allows the model to be tested using well-constrained experiments rather than poorly constrained gas-phase abundances in ISM regions. Our results indicate that inclusion of non-thermal reactions and suprathermal species allows for reproduction of low-temperature solid-phase photoprocessing that simulate interstellar ices within cold ($\sim$ 10 K) dense cores such as TMC-1. 

\end{abstract}



\section{Introduction} \label{sec:introduction}

While gas-phase and surface reactions on bare carbonaceous or silicate dust grains contribute to cosmic chemistry, the energetic processing of cosmic ices within dark, dense molecular clouds via photochemistry (initiated by non-ionizing radiation) and radiation chemistry (initiated by ionizing radiation) is thought to be the dominant mechanism for the interstellar synthesis of prebiotic molecules (see, for example, review: \citep{arumainayagam_extraterrestrial_2019}). Rate-equation based modeling treatments of UV-induced condensed-phase photochemistry have been moderately successful in reproducing the abundances of complex organic molecules (COMs) observed toward hot cores/corinos  \citep{shingledecker_simulating_2019,grassi_krome_2014,mcelroy_umist_2013,garrod_three-phase_2013}. However, recent detections of several COMs (e.g., methyl formate (\ce{HCOOCH3}) and dimethyl ether (\ce{CH3OCH3})) in cold ($\sim$ 10 K) dense cores \citep{vastel_origin_2014,taquet_chemical_2017,scibelli_prevalence_2020,bacmann_detection_2012,jimenez-serra_spatial_2016,oberg_disk_2010}, albeit in smaller abundance than in hot cores, have led to the search for alternative mechanisms for complex molecule production through cold or non-thermal mechanisms \citep{shingledecker_cosmic-ray-driven_2018,vasyunin_formation_2017}. A recent radiolysis-related computational study \citep{shingledecker_simulating_2019} has provided an explanation for the unprecedented observations of chemical synthesis at temperatures as low as 10 K in starless and prestellar cores. In this modified bulk-chemistry method involving radiolysis by cosmic rays, radicals produced within the ice are considered to be trapped and attempt to react with a neighbor with approximately every vibration. In the study described herein, we use this non-diffusive mechanism to revise the treatment of solid-phase photoprocesses in astrochemical models to account for the complex organics observed in cold cores. A recent study by \citet{jin_formation_2020} utilizes a non-diffusive rate-based model which demonstrates the dependence of COM production on non-diffusive reactions between radicals and ice species in cold astrochemical environments and achieves considerable success in reproducing observations toward prestellar core L1544. In contrast to that study, the model presented here incorporates (1) the detailed inclusion of photoionization and photoexcitation, and (2) explicit consideration of the production and reaction of electronically excited radicals (suprathermal species). 

One of the main processing mechanisms of ices in molecular clouds is radiation chemistry, which involves ionization and the production of copious numbers of low-energy ($<$ 15 eV) electrons, which are thought to be the dominant species involved in radiation chemistry (e.g.,\citep{arumainayagam_low-energy_2010}). Ionizing radiation present in this environment include MeV to TeV cosmic rays ($\sim$ 85\% H$^+$, $\sim$ 13\% He$^{2+}$, $\sim$ 1\% heavy bare nuclei, and $\sim$ 1\% electrons) and high-energy photons (e.g., vacuum ultraviolet photons with energies higher than $\sim$ 10 eV, extreme ultraviolet, X-ray, and $\gamma$-ray).  

Whereas high-energy photons contribute to radiation chemistry in dense molecular clouds, low-energy ($<$ 10 eV) photons (e.g., far (deep)-UV (4.1 – 6.2 eV)) initiate photochemistry, a process that does not involve direct ionization, but instead involves reactions of electronically-excited species. The UV interstellar radiation field, consisting of radiation from nearby stars, is extinguished by dust well before reaching the interior of dark, dense molecular clouds where prebiotic molecules are synthesized. However, a local secondary UV field exists to initiate photo-processing of dust grain ices \citep{prasad_uv_1983}. Cosmic rays excite gaseous molecular hydrogen, resulting in Lyman and Werner band emission with an estimated flux of $\sim$ 10$^3-$ 10$^4$ photons cm$^{-2}$ s$^{-1}$ \citep{gredel_cosmic-ray-induced_1989,cruz-diaz_vacuum-uv_2014,shen_cosmic_2004}. Although this spectral distribution includes high-energy (10 to 13.6 eV) photons, over half of the secondary UV field consists of low-energy ($<$ 10 eV) photons capable of photochemistry by exciting condensed species which may then react within the ice. Except during  high photon-flux laser experiments which may involve multi-photon processes, photochemistry is subject to the Bunsen-Roscoe law, which states that the photochemical yield is directly proportional to dose, irrespective of dose rate; this law allows for extrapolation from laboratory experiments to real astrochemical predictions, though even these low-flux experiments utilize fluxes much higher than those experienced by ice in dark, dense molecular clouds.  

Astrochemical models provide a critical link between the fundamental chemical information revealed by laboratory experiments and predictions and observations of chemical abundances in the interstellar medium. Most models utilize a rate-based approach due to the convenience and speed of that method, though Monte Carlo models have been used to more accurately simulate processes such as the catastrophic impact of high energy radiation or multi-layer interactions \citep{cuppen_grain_2017,oberg_photochemistry_2016}. Because all such simulations involving reaction networks and rate-equations are extremely sensitive to input parameters, these models generally become more successful as parameters are better constrained by laboratory experiments. Abundances of several COMs in hot cores/corinos are well reproduced by modern rate-equation-based computational models, which include a coupled gas-phase and grain-surface chemistry or three-phase (gas, surface, and bulk) chemistry \citep{cuppen_grain_2017}. 

Abundances of COMs in cold ($\sim$ 10 K) cores, however, are generally underpredicted. Most astrochemical models that include bulk-phase processes require thermal diffusion before reaction \citep{cuppen_grain_2017}; however, in the low-temperature conditions of starless and prestellar cores, this thermal motion within the bulk ice is not feasible. Several explanations for the formation of COMs in cold dense cores have been proposed, including: (1) photo-processing followed by reactive desorption \citep{watanabe_efficient_2002,chuang_production_2017,aikawa_molecular_2008,herbst_complex_2009,jin_formation_2020}, (2) gas-phase reactions \citep{balucani_formation_2015,codella_seeds_2020}, and (3) astrophysical-shock-catalyzed chemistry \citep{james_tracing_2020}, (4) methanol reactive desorption on CO-rich ices (within the catastrophic CO-freeze-out zone of pre-stellar cores), followed by gas-phase chemistry \citep{vasyunin_formation_2017}. The model presented herein instead assumes that reactive species are trapped within the bulk ice, but have the possibility of reacting with neighboring molecules during each vibration. This work is an extension of a previous study (modeling the physicochemical effects of astrochemical \ce{O2} and \ce{H2O} ice-analogue bombardment by energetic protons), which revealed the importance of considering fast non-thermal reactions in these systems \citep{shingledecker_simulating_2019}. In what follows, we apply this assumption to the case of photon irradiation of cosmic-ice analogues. The model, as utilized in this work, includes only photon-initiated ice processing, including both excitation and ionization events. Cations produced via ionization are assumed to quickly recombine with a secondary electron, resulting in the electronically excited parent molecule, which can then dissociate into electronically-excited products. Models such as ours are essential for interpreting planetary and interstellar ice data generated by past, upcoming, and ongoing NASA missions such as Spitzer, Stratospheric Observatory for Infrared Astronomy (SOFIA), and the James Webb Space Telescope (JWST).  

As a proof of concept, this new model is used to simulate two published laboratory studies that monitored the processing of \ce{O2} ice by $<$ 10.8 eV photons. Given that interstellar ice mantles are likely segregated into polar and apolar layers, it is useful to ``tune'' the model to simulate the photo-processing of a single species accurately, and then these species-specific models may be combined to simulate processing of the layers of more realistic cosmic ice analogues  \citep{tielens_interstellar_1991,pontoppidan_spatial_2006,oberg_quantification_2009,oberg_thespitzerice_2011}. The first study \citep{gerakines_ultraviolet_1996} employed a microwave-discharge hydrogen-flow lamp (MDHL) with a photon flux of $2.2\times10^{14}$ photons cm$^{-2}$ s$^{-1}$. The MDHL spectrum closely reproduces the calculated dark, dense molecular cloud secondary UV spectrum above 115 nm (below 10.8 eV). However, the fraction of Lyman alpha emission in a MDHL spectrum can change significantly based on the experimental settings such as microwave power and gas pressure \citep{ligterink_controlling_2015}. In \citet{gerakines_ultraviolet_1996}, oxygen ices ($\sim$ 100 nm in thickness) were deposited at $\sim$ 10 K inside a vacuum chamber 10$^{-7}$ Torr, mimicking conditions relevant to those of interstellar ices in cold cores. Two capping layers of argon precluded both contamination and significant desorption from the ices during photon irradiation. The oxygen ice was irradiated with photons for one hour corresponding to a maximum fluence of $7.9\times10^{17}$ photons cm$^{-2}$, corresponding to approximately a million years of secondary interstellar UV irradiation. Production of \ce{O3} during the irradiation was monitored via the 1043 cm$^{-1}$ IR feature of \ce{O3}. 

In the second experiment \citep{raut_photolysis_2011}, a pulsed ArF excimer laser (193 nm), defocused using a \ce{MgF2} lens, with a flux of $\sim2.3\times10^{14}$ photons cm$^{-2}$ s$^{-1}$, was used to irradiate \ce{O2} ices 80–84 nm thick. The use of 6.4 eV photons precludes radiation chemistry. The maximum fluence was $9.3\times10^{18}$ photons cm$^{-2}$. A vacuum chamber with a base pressure of 10$^{-9}$ Torr and an ice temperature of 22 K simulated interstellar-like conditions. The ozone column density as a function of fluence was monitored via the 1043 cm$^{-1}$ and 2109 cm$^{-1}$ IR features of \ce{O3}.  

The goal of the present work is to improve the current understanding of ice chemistry initiated by interstellar secondary UV radiation within dark, dense molecular clouds during the starless and prestellar stages well before the formation of protostars and planets. The successful reproduction of experimental results herein indicates that the inclusion of non-thermal reactions and suprathermal species will allow for more accurate modeling of interstellar photoprocessing of ices in cold cores. Additional tuning of the model for other species, such as water, will render the model suitable for predicting cold-core COM abundances attributable to photo-processing of mixed ices.  

\section{Methods} \label{sec:methods}

\subsection{Theory} \label{sec:theory}

As in \citet{shingledecker_general_2018}, the starting point of our proposed model is the assumption that the interaction between a UV photon and some target species, $A$, results in one of the following outcomes: 

\begin{equation}
A  \leadsto A^+ + e^- 
\tag{P1}
\label{p1}
\end{equation}

\begin{equation}
A  \leadsto A^+ + e^- \rightarrow A^* \rightarrow B^* + C^* 
\tag{P2}
\label{p2}
\end{equation}

\begin{equation}
A  \leadsto A^* \rightarrow B + C 
\tag{P3}
\label{p3}
\end{equation}

\begin{equation}
A  \leadsto A^* 
\tag{P4}
\label{p4}
\end{equation}

\noindent
Here, the curly arrow ($\leadsto$) represents the absorption of a photons, \textit{B} and \textit{C} are dissociation products, and * indicates an electronically excited (suprathermal) species. Of the four processes given above, \eqref{p1} and \eqref{p2} correspond to the photoionization of $A$ to the cation $A^+$— followed by the rapid recombination of the charged products in the case of \eqref{p2} — and are relevant in solids for $h\nu\gtrapprox10$ eV \citep{arumainayagam_extraterrestrial_2019}. Similarly, \eqref{p3} and \eqref{p4} account for photoexcitation to the excited state $A^*$, with \eqref{p3} leading to the photodissociation of $A$. One advantage of separating photoionization and photoexcitation is that the former can be enabled or disabled based on the energy of the incident photons. 

The rate coefficients of photoionization and dissociation processes, $k_\mathrm{photo}$, are usually calculated using 

\begin{equation}
    k_\mathrm{photo} = \int \sigma(\lambda)I(\lambda)d\lambda
    \label{kphotoNormal}
\end{equation}

\noindent
where here, $\sigma(\lambda)$ and $I(\lambda)$ are wavelength-dependent cross-section and photon flux, respectively. This formula can also be expressed as 

\begin{equation}
    k_\mathrm{photo} = \frac{\int \sigma(\lambda)I(\lambda)d\lambda}{\int I(\lambda)d\lambda}\int I(\lambda)d\lambda = \bar{\sigma}\Phi
    \label{kphotoExpanded}
\end{equation}

\noindent
where $\bar{\sigma}$ is the average cross-section, and $\Phi$ is the integrated photon flux. Following \citet{shingledecker_general_2018}, we can then express the rates of \eqref{p1} – \eqref{p4} in the following way: 

\begin{equation}
  k_\mathrm{P1} = P_\mathrm{e}\bar{\sigma}_\mathrm{ion}\Phi\delta
  \label{kp1}
\end{equation}

\begin{equation}
  k_\mathrm{P2} = (1 - P_\mathrm{e})\bar{\sigma}_\mathrm{ion}\Phi\delta
  \label{kp2}
\end{equation}

\begin{equation}
  k_\mathrm{P3} = P_\mathrm{dis}\bar{\sigma}_\mathrm{exc}\Phi\delta
  \label{kp3}
\end{equation}

\begin{equation}
  k_\mathrm{P4} = (1 - P_\mathrm{dis})\bar{\sigma}_\mathrm{exc}\Phi\delta.
  \label{kp4}
\end{equation}

\noindent
Here, $P_\mathrm{e}$ is the electron escape probability \citep{elkomoss_parention_1962}, which we assume as a first approximation is equal to zero. A more comprehensive model will need to relax this approximation to account for the effects of low-energy secondary electrons thought to be the primary agents of radiation chemistry. All ionized molecules are assumed to quickly recombine to form an excited molecule which will subsequently dissociate, react, or be quenched. Quenching by the surrounding ice is assumed to be the dominant relaxation mechanism rather than radiative relaxation, and the attempt frequency is used as the first-order rate constant for this process \citep{shingledecker_case_2019}. In reality, electronic excitations (excitons) may diffuse from the interior of the ice to the selvedge where they can drive the desorption of species into the gas \citep{thrower_highly_2011, marchione_efficient_2016-1}. $P_\mathrm{dis}$ is the dissociation probability, which is $\sim$ 1 in the gas, but in solids, we assume it to be 0.5 as a first approximation. This value was adjusted to account for spectral characteristics in later simulations. Dissociation products can recombine to reform the parent species, but this recombination is not assumed to occur preferentially to – or is calculated differently than - any other possible chemical reaction with other bulk species the fragments could undergo. The $\delta$ is a fitting factor that was introduced to account for assumptions of the model and absolute uncertainties in experimental data such as photon flux. More explicitly, $\delta$ is sensitive to, e.g., (a) the reactions and photoproducts included in the chemical network, (b) the associated branching fractions or cross sections, as well as (c) the methods for treating the underlying physical processes employed in the code. Thus, reasonable agreement between calculated and experimental data obtained assuming $\delta \approx 1$ for all photoprocesses would suggest that (a), (b), and (c) capture the salient features of a given system. Conversely, shortcomings in (a), (b) or (c) can be compensated for to some degree by adjusting $\delta$ values to yield best agreement with experimental results. For photoprocesses occurring in the bulk of optically thick ices, an extinction factor, $\epsilon$, can be included in Eqs. \eqref{kp1} – \eqref{kp4} to account for the reduced photon flux relative to the ice surface. Because the two experimental studies of interest used optically thin ices, the extinction factor was set equal to 1. 

\subsection{Model} \label{sec:model}

\begin{deluxetable*}{l|l|l|l}
\tabletypesize{\footnotesize}
\tablenum{1}
\tablecaption{Reactions comprising the chemical network used to model \ce{O2} photo-processing and subsequent chemistry. \label{tab:network}}
\tablewidth{0pt}
\tablehead{
\multicolumn{1}{c|}{Photon-Induced Reactions} & \multicolumn{3}{c}{Non-Photon-Induced Reactions} \\
}
\startdata
$\ce{O}  \xrightarrow{ionization} \ce{O}^+ + \ce{e}^-  \rightarrow  \ce{O}^*$ & $\ce{O3}^* + \ce{O3}  \rightarrow  \ce{O2} + \ce{O2} + \ce{O2}$ & $\ce{O2}^* + \ce{O}  \rightarrow  \ce{O3}^*$ & $\ce{O}^* + \ce{O3}^*  \rightarrow  \ce{O2} + \ce{O2}$ \\
$\ce{O2}  \xrightarrow{ionization}  \ce{O2}^+ + \ce{e}^-  \rightarrow \ce{O2}^*  \rightarrow  \ce{O}^* + \ce {O}^*$ & $\ce{O} + \ce{O}  \rightarrow  \ce{O2}^*$ & $\ce{O2}^* + \ce{O}  \rightarrow  \ce{O2} + \ce{O}$ & $\ce{O2}^* + \ce{O2}^*  \rightarrow  \ce{O2} + \ce{O2}$ \\
$\ce{O3}  \xrightarrow{ionization}  \ce{O3}^+ + \ce{e}^-  \rightarrow  \ce{O3}^*  \rightarrow  \ce{O2}^* + \ce{O}^*$ & $\ce{O}^* + \ce{O}  \rightarrow  \ce{O} + \ce{O}$ & $\ce{O2}^* + \ce{O2}  \rightarrow  \ce{O2} + \ce{O2}$ & $\ce{O2}^* + \ce{O3}^*  \rightarrow  \ce{O2} + \ce{O2} + \ce{O}$ \\
$\ce{O}  \xrightarrow{excitation}  \ce{O}^*$ & $\ce{O}^* + \ce{O}  \rightarrow  \ce{O2}^*$ & $\ce{O2}^* + \ce{O3}  \rightarrow  \ce{O2} + \ce{O2} + \ce{O}$ & $\ce{O3}^* + \ce{O3}^*  \rightarrow  \ce{O2} + \ce{O2} + \ce{O2}$ \\
$\ce{O2}  \xrightarrow{excitation}  \ce{O2}^*$ & $\ce{O}^* + \ce{O2}  \rightarrow  \ce{O3}^*$ & $\ce{O}^* + \ce{O}^*  \rightarrow  \ce{O} + \ce{O}$ & $\ce{O} + \ce{O3}^*  \rightarrow  \ce{O2} + \ce{O2}$ \\
$\ce{O3}  \xrightarrow{excitation}  \ce{O3}^*$ & $\ce{O}^* + \ce{O2}  \rightarrow  \ce{O} + \ce{O2}$ & $\ce{O}^* + \ce{O2}^*  \rightarrow  \ce{O3}^*$ & $\ce{O2} + \ce{O3}^*  \rightarrow  \ce{O2} + \ce{O2} + \ce{O}$ \\
$\ce{O2}  \xrightarrow{excitation}  \ce{O2}^*  \rightarrow  \ce{O} + \ce{O}$ & $\ce{O} + \ce{O3}  \rightarrow  \ce{O2} + \ce{O2}$ & $\ce{O}^* + \ce{O2}^*  \rightarrow  \ce{O} + \ce{O2}$ & $\ce{O} + \ce{O2}  \rightarrow  \ce{O3}^*$ \\
$\ce{O3}  \xrightarrow{excitation}  \ce{O3}^*  \rightarrow  \ce{O2} + \ce{O}$ & $\ce{O}^* + \ce{O3}  \rightarrow  \ce{O2} + \ce{O2}$ & \\
\enddata
\end{deluxetable*}

In this work, we have utilized the \texttt{MONACO} model \citep{vasyunin_formation_2017}, previously modified by us, to simulate ice radiation chemistry experiments \citep{shingledecker_simulating_2019}. This code, written in Fortran 90, solves a system of coupled differential equations describing the evolution of the abundance of each species in our network. Unlike comparable astrochemical models, the model described herein accounts for electronically-excited suprathermal species produced during photo-processing of ices. Table \ref{tab:network} presents all photon-induced and non-photon-induced (reactions involving products of the initial photo-processing) reactions included in the model network for \ce{O2}. Reactions occurring in the selvedge, considered to be the top four monolayers of the ice \citep{vasyunin_unified_2013},  are assumed to occur via the Langmuir-Hinshelwood mechanism, and rate coefficients are calculated using the standard formula for diffusive processes. For reactions in the bulk, rate coefficients are calculated using the non-diffusive formula of \citet{shingledecker_simulating_2019},

\begin{equation}
  k_\mathrm{fast} = 
  f_\mathrm{br}
  \left[
  \frac{
  \nu_0^A + 
  \nu_0^B
  }
  {
  {N_\mathrm{bulk}}
  }
  \right]
  \mathrm{exp}
  \left(-
  \frac 
  {E_\mathrm{act}^{AB}}
  {T_\mathrm{ice}}
  \right),
  \label{ksup}
\end{equation}

\noindent
where  $f_\mathrm{br}$ is the branching fraction, $T_\mathrm{ice}$ is the ice temperature, $E_\mathrm{act}^{AB}$ is the activation energy in Kelvins for reaction, $N_\mathrm{bulk}$ is the total number of bulk species in the simulated ice, and $\nu^A_0$ is the characteristic (hereafter, trial) vibrational frequency \citep{herbst_chemistry_2008}. 

In our model, species in the selvedge can, in principle, desorb both thermally as well as following exothermic association reactions, with the latter being treated by the method of \citet{garrod_new_2008} with a standard efficiency of 1\%. To mimic the Gerakines experiments where such desorption would be inhibited due to the presence of a capping noble gas layer, thermal, chemical, and photodesorption processes were disabled in our models. The ices studied by Raut and coworkers lacked such a noble gas cap, and thus some amount of desorption would have occurred during the course of the experiment. However, given the fairly low temperature (22 K) of the bulk ice, we assume that thermal desorption is negligible over the timescale of the experiment. Moreover, photodesorption for the Raut et al. experiments was also disabled, since the desorption rate is not well constrained \citep{fayolle_wavelength-dependent_2013,bulak_novel_2020} and our focus for this study was, in any case, the chemistry occurring \textit{within} the bulk ice. Chemical desorption is included for the simulation of the Raut et al. data but was found to have a negligible impact on the bulk chemistry we describe in detail below. 

Rate constants for photon-induced reactions are dependent on the average cross-sections ($\bar{\sigma}$), photon flux ($\Phi$), and the fitting factor ($\delta$) as described in \S\ref{sec:theory}. Each product of photoionization or photoexcitation is treated as being trapped in a cage of neighboring bulk species molecules; reactions involving suprathermal species are assumed to be barrierless. Non-photon-induced reactions are assumed to occur non-diffusively, with rates of reaction between any two species being proportional to their abundances in the ice \citep{shingledecker_simulating_2019}.  

For the pure \ce{O2} ice, we use the chemical network (Table \ref{tab:network}) initially described in \citet{shingledecker_new_2017} and used in the microscopic Monte Carlo model, \texttt{CIRIS}, and later modified for use in rate-based kinetic codes in \citet{shingledecker_simulating_2019}. The choice of \ce{O2} as the bulk species for this initial test of the model is appropriate given the relative simplicity of the products and subsequent possible reactions, especially compared to species such as water or methanol. The selvedge, which comprises the chemically distinct region near the top of the mantle, is considered to be the top four monolayers of the ice \citep{vasyunin_unified_2013}. 

Parameters relevant to the simulation of laboratory experiments include ice thickness, photon fluence (photon flux multiplied by irradiation time), and photon energy; these values were obtained directly from the manuscripts of the experiments chosen for simulation. The trial frequency, $\nu$, parameterizes the vibrational frequency of a molecule, used as the pre-exponential factor in calculating bulk rate coefficients. The model assumes that with every vibration, there is a probability that a molecule will react with a neighboring molecule (Eq. \eqref{ksup}). For all simulations, the vibrational frequency was set to $1\times10^{15}$ s$^{-1}$, which is reasonable, assuming RRKM theory. Increasing the value by orders of magnitude has negligible impact on model simulations, while reducing the value below $1\times10^{15}$ s$^{-1}$ resulted in significant deviations from experimental data.

Specific to each photoprocess included in the chemical network are cross-sections and the $\delta$ fitting factor. Branching ratios for reactions with more than one product channel were assumed to occur with equal probability in early simulations, and later adjusted to match spectral characteristics. Cross-sections were obtained from various sources, as detailed below. 

\section{Results and Discussion} \label{sec:results}

\begin{deluxetable*}{lcllcc}
\tablenum{2}
\tablecaption{Calculated average cross-sections and $\delta$-values for pure \ce{O2} ice irradiated by a MDHL and ArF laser. \label{tab:sigmadelta}}
\tablewidth{0pt}
\tablehead{
\colhead{Process} & \colhead{Type} & \colhead{$\bar{\sigma}_\mathrm{MDHL}$ (cm$^2$)} & \colhead{$\bar{\sigma}_\mathrm{ArF}$ (cm$^2$)} & \colhead{$\delta_\mathrm{MDHL}$} & \colhead{$\delta_\mathrm{ArF}$} \\
}
\startdata
$\ce{O} + h\nu \rightarrow \ce{O}^*$ & \eqref{p2} & 0 & 0 & 1.0 & 1.0 \\
$\ce{O} + h\nu \rightarrow \ce{O}^*$ & \eqref{p4} & 0 & 0 & 1.0 & 1.0 \\
$\ce{O2} + h\nu \rightarrow \ce{O}^* + \ce{O}^*$ & \eqref{p2} & $3.86\times10^{-20}$ & $0$ & 1.0 & 1.0 \\
$\ce{O2} + h\nu \rightarrow \ce{O} + \ce{O}$ & \eqref{p3} & $2.13\times10^{-18}$ & $2.10\times10^{-19}$ & 2.3 & 1.9 - 2.2 \\
$\ce{O2} + h\nu \rightarrow \ce{O2}^*$ & \eqref{p4} & $2.13\times10^{-18}$ & $2.10\times10^{-19}$ & 1.0 & 0.25 - 0.35 \\
$\ce{O3} + h\nu \rightarrow \ce{O2}^* + \ce{O}^*$ & \eqref{p2} & 0 & 0 & 1.0 & 1.0 \\
$\ce{O3} + h\nu \rightarrow \ce{O2} + \ce{O}$ & \eqref{p3} & $5.60\times10^{-18}$ & $2.15\times10^{-18}$ & 1.0 & 0.25 - 0.35 \\
$\ce{O3} + h\nu \rightarrow \ce{O3}^*$ & \eqref{p4} & $5.60\times10^{-18}$ & $2.15\times10^{-18}$ & 1.0 & 0.25 - 0.35 \\
\enddata
\tablecomments{the values of $\delta_\mathrm{MDHL}$, were used to produce Figure \ref{fig:gerakinesExp}. For $\delta_\mathrm{ArF}$, ranges of values are shown which were found to yield agreement with the data within experimental error.}
\end{deluxetable*}

We directly tested the validity of our new method by replicating experimental data of \ce{O2} ice irradiation with UV photons ($< 10.8$ eV). In contrast to simulations of poorly constrained gas-phase abundances in ISM regions, this method allows the model to be tested using well-constrained experiments. Simulated pure oxygen ice experiments shared the use of interstellar-like temperatures and pressures. The results of these simulations are described below. 

\subsection{Microwave discharge hydrogen flow lamp source}

\begin{figure}[htb]
    \centering
    \includegraphics[width=\textwidth]{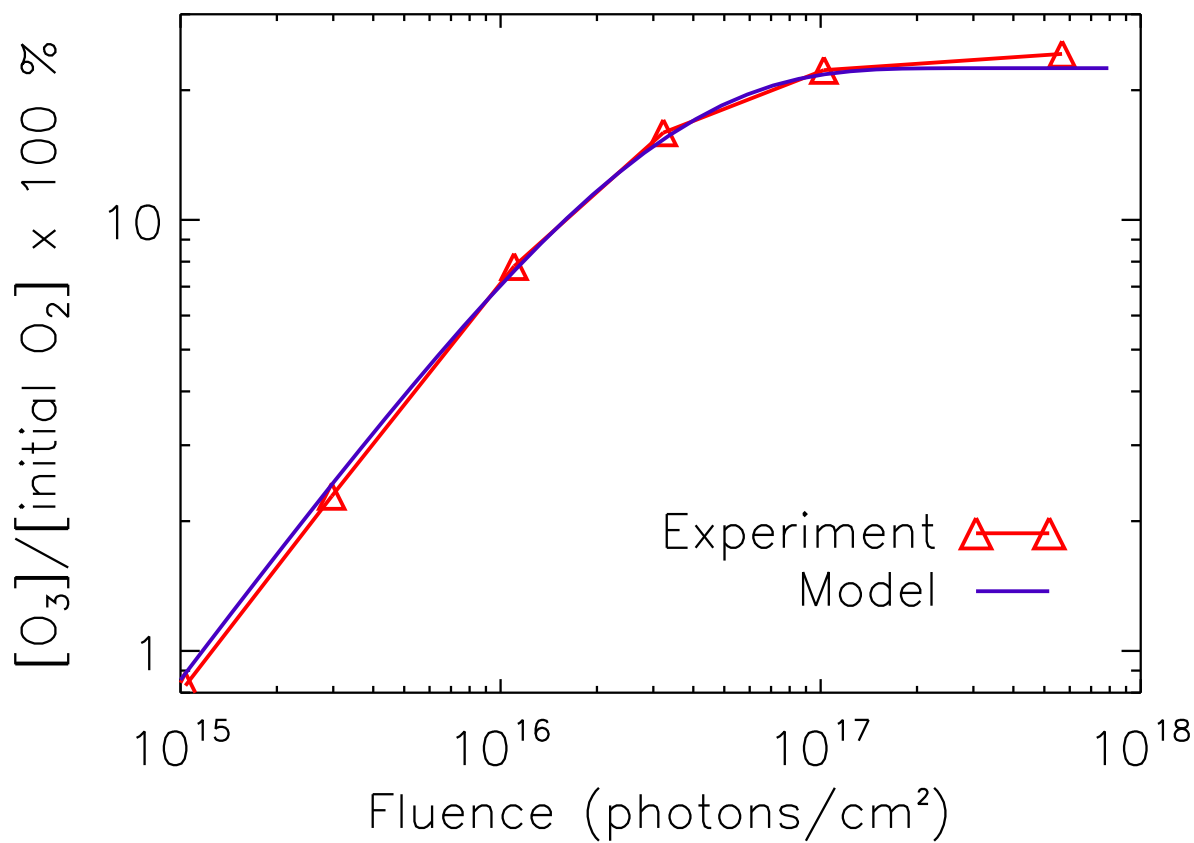}
    \caption{Calculated abundances of \ce{O3} vs. photon fluence in a UV-irradiated pure \ce{O2} ice (shown in blue), with corresponding experimental data from \citet{gerakines_ultraviolet_1996} (shown in red). The values shown in Table \ref{tab:sigmadelta} were used for branching ratios and cross sections. No error bars were reported in this study}
    \label{fig:gerakinesExp}
\end{figure}

Previous work by \citet{gerakines_ultraviolet_1996} provides excellent data with which to quantify the validity of our approach. Their experiments on pure \ce{O2} were carried out at 10 K. They utilized a microwave discharge hydrogen flow lamp (MDHL), two layers of inert argon, film thicknesses on the order of 0.1 $\mu$m, photon fluxes of $\sim10^{14}$ photons cm$^{-2}$ s$^{-1}$, and an irradiation time of $\sim$ 1 hr \citep{gerakines_ultraviolet_1996,jenniskens_carbon_1993}. The \ce{O3} production curve given in figure 8 of Gerakines et al. was digitized for comparison to the model output of \ce{O3} abundance (Fig. \ref{fig:gerakinesExp}). Listed in Table \ref{tab:sigmadelta} are the effective cross sections, $\bar{\sigma}$, for this experiment. To calculate the effective solid-phase cross section for $\ce{O2}$ photoabsorption, first a reported spectrum of solid-phase \ce{O2} absorption (but not cross section) as a function of wavelength was digitized \citep{lu_absorption_2008}. Next, this absorption data was scaled to the digitized solid phase cross section data of \citet{mason_vuv_2006} in order to obtain cross section values over a broader wavelength range corresponding to the spectrum of a MDHL. The spectrum of a MDHL  \citep{jenniskens_carbon_1993} was digitized, and intensity and cross section were multiplied together at each wavelength. Finally, the product was integrated over all wavelengths and divided by total flux (Eq. \eqref{kphotoExpanded}). The effective solid-phase $\ce{O3}$ photoabsorption cross section was calculated by first scaling gas-phase absorption data from \citet{sivaraman_vacuum_2014} to gas-phase cross section data from the Leiden database \citep{heays_photodissociation_2017}. Solid-phase absorption data taken in the same laboratory \citep{sivaraman_vacuum_2014} were multiplied by the same scaling factor. The resulting solid-phase photoabsorption cross section data as a function of wavelength was used to calculate effective solid-phase cross section. To obtain cross section values for the excited-state reaction path  (e.g., \eqref{p4}) and the dissociation path (e.g., \eqref{p3}), total photoabsorption cross-sections were multiplied by the corresponding branching ratio. Average solid-phase cross sections for $\ce{O2}$ and $\ce{O3}$ photoionization were obtained by using gas phase cross section data from the Leiden database but shifting all data points by 1.5 eV for application to condensed species \citep{kahn_fermi_2015, yu_ups_1975}.
After inserting the experimental parameters of ice thickness, photon flux, photon-irradiation time, and reaction cross sections, the $\delta$ fitting factors given in Table \ref{tab:sigmadelta} were obtained by manually adjusting to maximize the agreement with experimental data. The optimized steady-state \ce{O3} abundance agrees with the experimental value to within 5\%. 

\begin{figure}[htb]
    \centering
    \includegraphics[width=\textwidth]{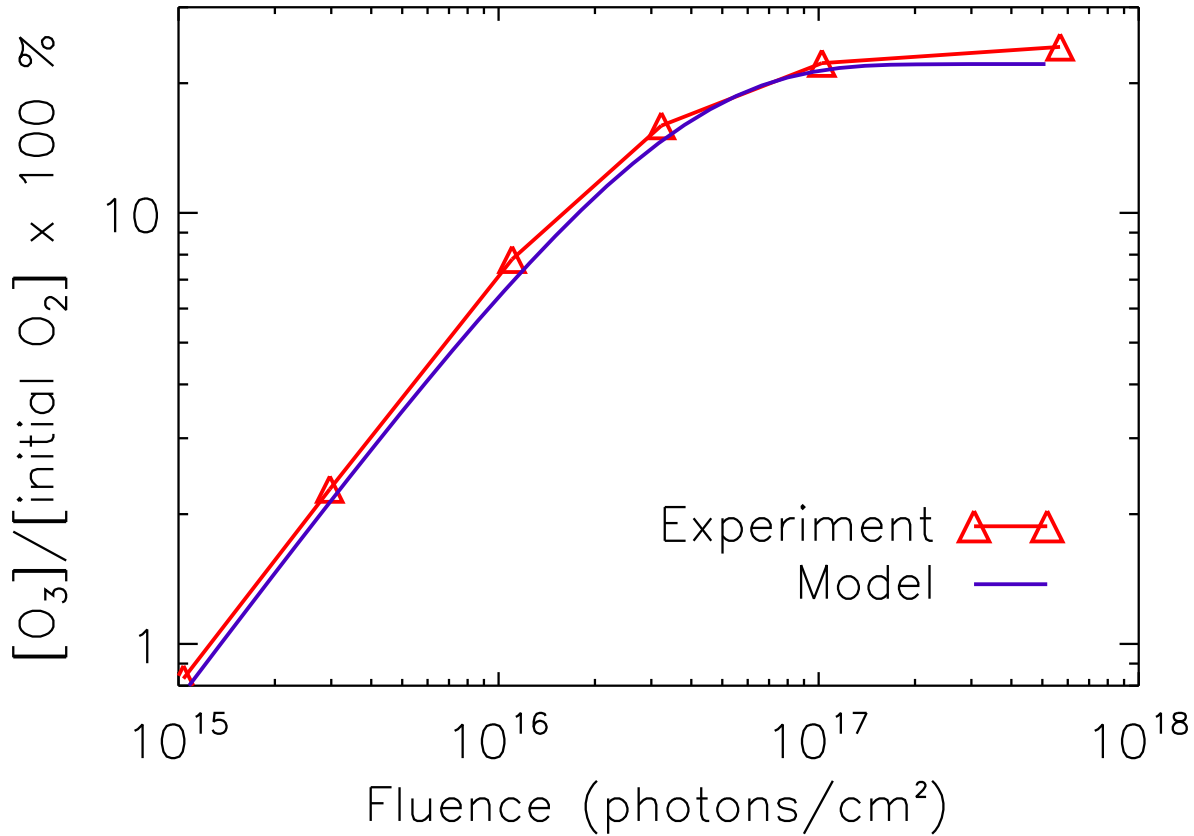}
    \caption{Calculated abundances of \ce{O3} vs. photon fluence in a UV-photodissociated pure \ce{O2} ice (shown in blue), with corresponding experimental data from \citet{gerakines_ultraviolet_1996} (shown in red). For this simulation, the branching ratio of $\ce{O2}$ photodissocation to $\ce{O2}$ photoexcitement was set to 0.99, and all $\delta$-values were set to 1.0.}
    \label{fig:gerakinesDelta1}
\end{figure}

Only the $\delta$ fitting factor for $\ce{O2}$ photodissociation varied from 1.0 for agreement with the experimental data. Given the broad absorption cross-section peak even in the solid-phase spectrum and the high probability of dissociation following photoabsorption for gaseous $\ce{O2}$, a simulation was run with a branching ratio of 99:1 for $\ce{O2}$ dissociation to $\ce{O2}$ excited-state reaction, the results of which are shown in Fig. \ref{fig:gerakinesDelta1}. All $\delta$ values could then be set to 1.0 for similar agreement to experimental data as when the dissociation and excited-state reactions were assumed to be equally likely but a $\delta$ value of 2.3 was required for the dissociation channel. Thus, the original deviation of $\delta$ from 1.0 was necessary to account for the  high \ce{O2} photodissociation probability when it was not otherwise included in the model.

\subsection{Pulsed 193 nm ArF excimer laser source}

\begin{figure}[htb]
    \centering
    \includegraphics[width=\textwidth]{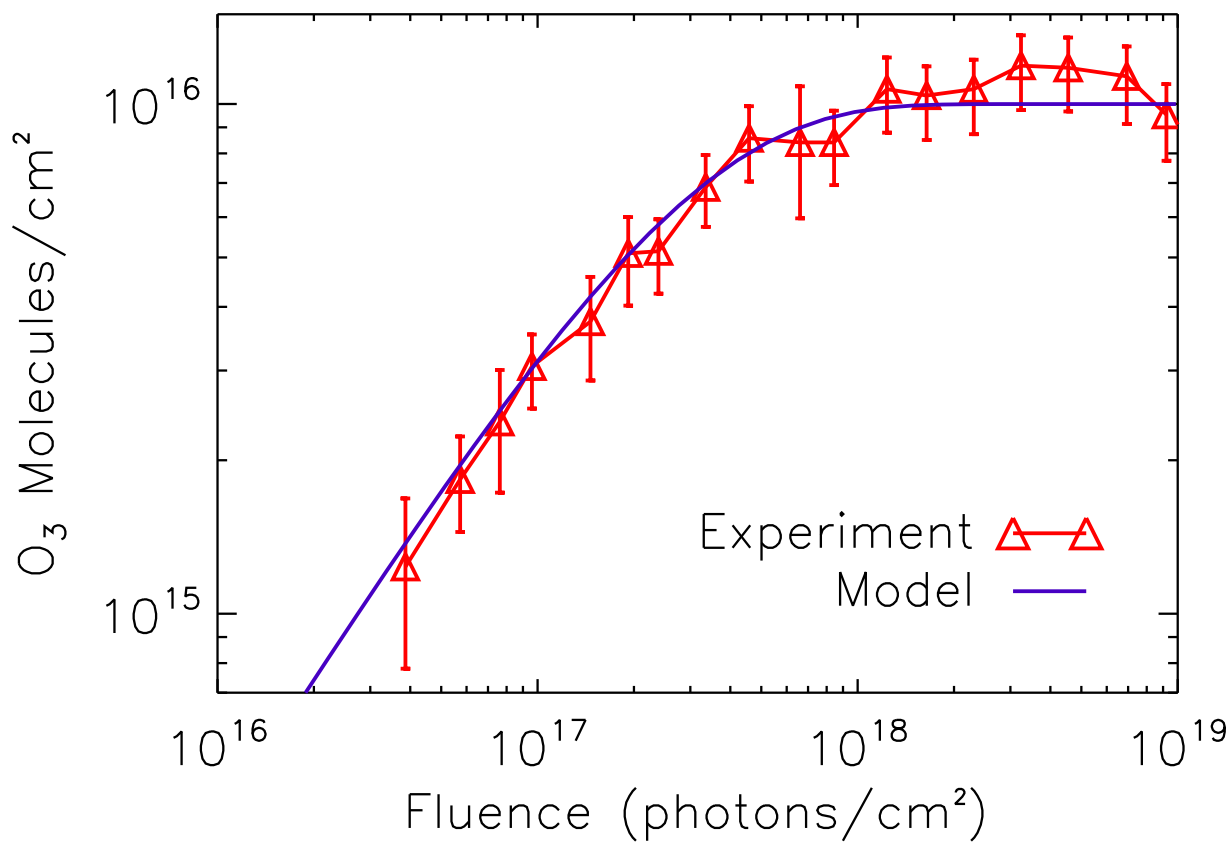}
    \caption{Calculated abundances of \ce{O3} vs. photon fluence in a UV-irradiated pure \ce{O2} ice (shown in blue), with corresponding experimental data from   \citet{raut_photolysis_2011} (shown in red). $\delta$-values used in this simulation: $\ce{O2} + h\nu \rightarrow \ce{O} + \ce{O}$ $[\delta = 2.0]$; $\ce{O2} + h\nu \rightarrow \ce{O2}^*$ $[\delta = 0.3]$; $\ce{O3} + h\nu \rightarrow \ce{O2} + \ce{O}$ $[\delta = 0.3]$; $\ce{O3} + h\nu \rightarrow \ce{O3}^*$ $[\delta = 0.3]$; $\delta = 1.0$ for all other processes.}
    \label{fig:RautExp}
\end{figure}

A different condensed-phase \ce{O2} experiment \citep{raut_photolysis_2011}, which utilized a pulsed laser UV source, was also simulated to test the validity of our model. These experiments were conducted at 22 K with a film thickness of $\sim$ 80 nm, a photon flux averaging $\sim10^{14}$ photons cm$^{-2}$ s$^{-1}$, and a total fluence of $\sim 10^{19}$ photons cm$^{-2}$ with 193 nm (6.4 eV) photons. The \ce{O3} abundance data in Figure 5 of Raut et al. was digitized for comparison to the model output. Because the \ce{O3} abundance was reported as column density, the model output was scaled to account for the thickness of the ice. In this case, cross-sections were provided in the experiment manuscript: the solid phase 193 nm photoabsorption cross section for \ce{O2} and \ce{O3} are reported as $4.2\times10^{-19}$ cm$^2$ and $4.3\times10^{-19}$ cm$^2$, respectively. To confirm these reported cross sections, two independently reported spectra of solid state \ce{O2} absorption (but not cross-section) as a function of wavelength were digitized \citep{cruz-diaz_vacuum-uv_2014,lu_absorption_2008}. Each included data at 193 nm. Next, a published spectrum of solid-phase \ce{O2} cross section as a function of wavelength was also digitized \citep{mason_vuv_2006}; this spectrum did not include data at 193 nm. The absorption spectra were then scaled to the cross-section spectrum. The values found in the scaled data at 193 nm matched the cross-section values used by Raut et al. to within 25\%.

Since Raut et al. included error values in their experimental data, a python script was used to find optimized $\delta$ fitting factors which would maximize agreement with experimental data. ``Maximal agreement'' was considered as any model output which fell entirely within upper and lower error. Our script iteratively ran the model over a range of $\delta$ values for each process and indicated which combinations of $\delta$ values resulted in model outputs which agreed with the data within experimental error; these are given in Table \ref{tab:sigmadelta}. Because there is a range of outputs which may fall within upper and lower error, there are correspondingly ranges of $\delta$ values for the most influential processes. Simulation results using $\delta$ values within this range are shown in Fig. \ref{fig:RautExp}.

As displayed in Table \ref{tab:sigmadelta}, the $\delta$ fitting factors, although close to unity, vary somewhat between the simulations of the two experiments. When branching ratios are adjusted from the initial assumption of equal likelihood to more realistic values, all fitting factors could be set to 1.0 for the Gerakines experiment. As noted in \S2.1, the $\delta$ fitting factors should be interpreted as effectively accounting for other factors (e.g., absolute uncertainties in the experimental data) not explicitly considered in the code. The fact that all $\delta$ values given in Table \ref{tab:sigmadelta} are close to unity indicates that the overall contribution of such unknown effects is likely small and are unlikely to be significant sources of uncertainly in astrophysical simulations. 

In our models, it was found that our \ce{O3} abundances were most sensitive to variations in the $\delta$ value for the $\mathrm{O_2} \leadsto \mathrm{2O}$ process, thereby revealing the importance of the \ce{O + O2} reaction on the overall abundance of ozone. This finding reveals another useful role for the $\delta$ values, namely, that of highlighting key reactions for a given species based on how sensitive the calculated abundance is to variations in the assumed values of $\delta$.

While the current model provides reasonable agreement with the findings of the considered experiments, a number of areas could be addressed in future studies that could increase both agreement with empirical data as well as the underlying physical realism of the simulation. As mentioned, the focus of this work has been on processes occurring in the mantles of thin ice films similar to those that coat interstellar dust grains, all of which were optically thin, and thus, we have not considered the effects of extinction that would be of particular importance in optically thick ices. To investigate this effect in more detail, a multi-layer model which more explicitly treats the vertical structure of ice mantles, such as the macroscopic Monte Carlo code described in \citet{vasyunin_unified_2013}, would be more appropriate.

Moreover, given the focus of this study on bulk chemistry, we have not considered photodesorption processes occurring in the top several monolayers of the ice. Current values used in models for these kinds of processes are not well constrained, however, our method of directly simulating laboratory experiments using astrochemical codes represents a promising means by which suitable values could be obtained.

Additionally, it is known that the absorption of photons of different energies will change both the efficiency and products of photo-processes \citep{fayolle_wavelength-dependent_2013,fayolle_co_2011,fillion_wavelength_2014}. Absent theoretical/experimental cross-sections for photoprocesses as a function of energy, we have as a first approximation assumed that, for example, the photodissociation of \ce{O2} produces with equal probability 2O$^*$ and 2O, but not O$^*$ + O. 

Finally, because subionization UV photons dominate the MDHL lamp/ArF laser UV, it is likely that the role of low-energy electrons is not significant in this study.  To simulate the effects of secondary UV radiation within dark, dense molecular clouds, this model must be modified to include secondary low-energy electron-induced process such as dissociative electron attachment that can occur at electron energies almost as low as 0 eV. 

\section{Conclusions} \label{sec:conclusions}

We have simulated the $<$ 10.8 eV UV photodissociation of solid \ce{O2} at 10-22 K by a microwave-discharge hydrogen flow lamp and an ArF excimer laser using a rate-based model. Our methodology incorporates: (a) non-diffusive bulk reactions for radicals and other reactive species and (b) a new theoretical method for simulating photoprocesses which, for the first time, distinguishes between photoexcitation and photoionization. We explicitly account for the production and reactivity of short-lived suprathermal photoproducts. In contrast to previous condensed phase astrochemical model calculations that have used gas-phase photoabsorption cross sections, we have employed solid-phase cross sections in our calculations. This method allows the model to be tested using well-constrained experiments rather than poorly constrained gas-phase abundances in regions of the ISM. The semi-quantitative agreement of the model with experimental \ce{O3}  abundances obtained in two different laboratories indicates that the methodology is promising for simulating interstellar ice photoprocessing. This new computational method,  focusing on non-diffusive reactions for radicals and suprathermal species, results in improved agreement with experimental data compared to techniques that rely on bulk thermal radical diffusion, an unlikely mechanism at the exceedingly low temperatures of cold cores. Ultimately it would be fruitful to incorporate these types of rate-based photoprocessing calculations into models that account for atom addition, gas-phase reactions, and cosmic-ray bombardment. Such models, together with observations and laboratory simulations, are necessary for a fundamental understanding of interstellar chemistry which is the likely source of prebiotic molecules in the universe.

\acknowledgments

C.N.S. thanks the Alexander von Humboldt Stiftung/Foundation for their generous support. E.M. gratefully acknowledges funding from the Arnold and Mabel Beckman Foundation. The Massachusetts Space Grant Consortium supported the work of MF. CRA's work was supported by grants from the National Science Foundation (NSF grant number CHE-1955215), Wellesley College (Faculty Awards and Brachman Hoffman small grants). Work by A.I.V. was supported by the Russian Ministry of Science and Higher Education, Project FEUZ-2020-0038

\software{MONACO - \citet{vasyunin_formation_2017}, MATLAB}

\bibliography{references,chrisa}
\bibliographystyle{aasjournal}



\end{document}